\documentclass[a4paper,fleqn,usenatbib]{mnras}

\usepackage{mathptmx}
\usepackage{txfonts}

\usepackage[T1]{fontenc}
\usepackage{ae,aecompl}

\usepackage{graphicx}	
\usepackage{amssymb}	
\usepackage{longtable}

\newcommand{\jgrl}{Geophysical Research Letters}
\newcommand{\jgrp}{Journal of Geophysical Research: Planets}
\newcommand{\jgra}{Journal of Geophysical Research: Atmospheres}
\newcommand{\jgrs}{Journal of Geophysical Research: Space Physics}
\newcommand{\jpca}{Journal of Physical Chemistry A}

\title[O($^3P$)+CO$_2$ scattering cross sections]{O($^3P$)+CO$_2$ scattering cross sections at superthermal collision energies for planetary aeronomy}

\author[M. Gacesa et al.]{
Marko Gacesa,$^{1,2}$\thanks{E-mail: marko.gacesa@nasa.gov}
R. J. Lillis,$^{3}$
and K. J. Zahnle$^{2}$
\\
$^{1}$Bay Area Environmental Research Institute, Moffett Field, California, USA \\
$^{2}$Space Science Division, NASA Ames Research Center, MS 245-3, Moffett Field, California, USA\\
$^{3}$Space Sciences Laboratory, University of California, Berkeley, California, USA
}

\date{Accepted XXX. Received YYY; in original form ZZZ}

\pubyear{2019}

\begin{document}
\label{firstpage}
\pagerange{\pageref{firstpage}--\pageref{lastpage}}
\maketitle

\begin{abstract}
We report new elastic and inelastic cross sections for O($^3P$)+CO$_2$ scattering at collision energies from 0.03 to 5 eV, of major importance to O escape from Mars, Venus, and CO$_2$-rich atmospheres. The cross sections were calculated from first principles using three newly constructed ab-initio potential energy surfaces correlating to the lowest energy asymptote of the complex. The surfaces were restricted to a planar geometry with the CO$_2$ molecule assumed to be in linear configuration fixed at equilibrium. Quantum-mechanical coupled-channel formalism with a large basis set was used to compute state-to-state integral and differential cross sections for elastic and inelastic O($^3P$)+CO$_2$ scattering between all pairs of rotational states of CO$_2$ molecule. The elastic cross sections are 35\% lower at 0.5 eV and more than 50\% lower at 4+ eV than values commonly used in studies of processes in upper and middle planetary atmospheres of Mars, Earth, Venus, and CO$_2$-rich planets. Momentum transfer cross sections, of interest for energy transport, were found to be proportionally lower than predicted by mass-scaling.
\end{abstract}

\begin{keywords}
scattering -- molecular processes -- planets and satellites: atmospheres -- planets and satellites: terrestrial planets -- planetary nebulae: general
\end{keywords}



\section{Introduction}

Understanding oxygen escape to space is a key to understanding evolution of the martian atmosphere. To first approximation, atomic hydrogen and oxygen escape from Mars are linked together. At present, H and O escape are directly related to water escape, as there is no pronounced evidence of excess of either species accumulating in the atmosphere. The global time-integrated rate of H escape is, therefore, equal to twice the global time-integrated O escape rate. 
While hydrogen escapes easily from Mars, mainly through thermal Jeans escape mechanism, atomic O escapes with difficulty due to its high mass \citep{1976AREPS...4..265H}. As a result, it is the O escape rate that is limiting and its escape mechanism determines the loss of water and, consequently, plays the biggest role in determining the atmospheric evolution on Mars \citep{1976Icar...28..231L,1976AREPS...4..265H,2003IJAsB...2..195L,2008JGRE..11311004Z}.
Curiously, the same 2:1 ratio is inferred in H-to-O escape rates from Venus \citep{1975Icar...24..148L}, suggesting principal importance of O escape at Venus, and for Venusian atmospheric evolution. 

The mechanism driving the O escape from Mars is largely photochemical \citep{1988GeoRL..15..433N,2018Icar..315..146J}. In its upper atmosphere, photodissociation of molecular O$_2$, O$_3$, CO$_{2}$, and CO by solar ultraviolet (UV) photons produces superthermal atomic oxygen in the ground state, O($^3P$), as well as electronically excited O($^1D$) and O($^1S$). 
Dissociative recombination of O$_2^+$ molecular ions with electrons preferentially produces superthermal O($^3P$), with kinetic energies up to 3.5 eV, capable of overcoming Mars' gravitational potential and escaping to space \citep{2009Icar..204..527F}. The photochemical escape appears to be the dominant non-thermal escape mechanism presently active on Mars \citep{2017JGRA..122.3815L,2017JGRA..122.1102C}.

Superthermal O atoms also drive escape of other atmospheric species, such as Ar, He, H$_2$, OH and H$_2$O, through kinetic energy transfer in collisions, contributing to additional non-thermal atmospheric loss \citep{doi:10.1029/2019GL082192}. 
The non-thermal atmospheric escape rates strongly depend on the attenuation of the hot O flux in the Martian upper atmosphere, which mainly depends on O($^3P$)+CO$_2$ scattering properties at superthermal energies.

Owing to the fact that CO$_2$ is a greenhouse gas as well as an important coolant (via emission at 15 $\mu$m line) in the middle and upper atmosphere of Earth and terrestrial planets in general \citep{sharma1990role}, the O($^3P$)+CO$_2$ scattering has been a subject of numerous investigations.
Early theoretical work focused mainly on energy transfer to and from specific vibrational states \citep{1974JChPh..60.2913B,1983CP.....80..213B,schatz1981quasiclassical,1982CPL....88..553M,1984CP.....87...63G}. They found that the energy transfer could be described well using impulse approximation in the superthermal regime, similar to collisions of CO$_2$ with noble gas atoms \citep{suzukawa1978quasiclassical}.
More recent studies covered the collision energies corresponding to temperatures 150 K $\leqslant T \leqslant 550$ K, of interest in terrestrial atmosphere, and focused on understanding collisional de-excitation (quenching) of vibrationally excited CO$_2$ by the O atoms, namely O($^3P$)+CO$_2$(0$1^1$0) $\rightarrow$ O($^3P$) + CO$_2$(0$0^0$0), as well as energy transfer between internal rotational and vibrational degrees of freedom \citep{2006JGRA..111.9303C,2006JChPh.124p4302D,2007MolPh.105.1171D,2012JGRA..117.4310C,2012ACP....12.9013F,2015CPL...638..149C}.

\citet{2012JPCA..116...64Y} experimentally and theoretically investigated O($^3P$)+CO$_2$ scattering at superthermal energies with isotopically labeled $^{12}$C$^{18}$O$_2$. In a crossed molecular beam experiment at 4.28 eV (equal to 98.8 kcal/mol), they measured elastic and inelastic scattering of $^{16}$O($^3P$)+$^{12}$C$^{16}$O$_2$, as well as rate constants for two reactions: isotope exchange and O atom abstraction. They also performed quasi-classical trajectory (QCT) calculations for collision energies from 1 eV to 6.5 eV. 
However, the study of \citet{2012JPCA..116...64Y} was focused on understanding the role of transient molecular states in reactive scattering processes, in particular the oxygen isotope exchange and its role in atomic oxygen cycling via CO$_2$ in the Earth's atmosphere, rather than on constructing detailed elastic and inelastic scattering cross sections at superthermal energies for purposes of planetary aeronomy.

In the absence of targeted studies, in planetary aeronomy O($^3P$)+CO$_2$ scattering cross sections are commonly estimated by mass-scaling from systems whose scattering properties at superthermal energies are known better, including O+O \citep{2000JGR...10524899K}, O+N$_2$ \citep{1998JGR...10323393B},  Ar+H$_2$ \citep{2005JChPh.122b4304U}, and O+H$_2$ \citep{2014JChPh.141p4324G}, or from general properties of atom-molecule cross sections at high collision energies \citep{2014ApJ...790...98L}.
For example, quantifying photochemical escape of oxygen from Mars \citep{2007JGRE..112.7001C,2009Icar..204..527F} and interpreting data from Mars Atmosphere and Volatile EvolutioN (MAVEN) mission \citep{2015JGRE..120.1880L,2015GeoRL..42.9015L,2015SSRv..195..357L,2017JGRA..122.3815L,2017JGRE..122.2401L,2017JGRE..122.1321A} and collisionally-induced escape processes driven by superthermal O atoms \citep{2011GeoRL..3802203B,2013oepa.book.....L,2012GeoRL..3910203G,2014ApJ...790...98L,2017Icar..284...90G,2017SoSyR..51..249S}, relied on mass-scaled cross sections specifically adapted for planetary aeronomy \citep{2014Icar..228..375F,2018Icar..300..411F}. 
These studies recognized that accurate O($^3P$)+CO$_2$ cross sections are needed to reduced the uncertainties in present day atmospheric escape from Mars and estimates of its primordial water inventory \citep{2017JGRA..122.3815L,2017JGRA..122.1102C,2018Icar..315..146J}.

Here, we present the results of a theoretical study of O($^3P$) + CO$_2(j)$ $\rightarrow$ O($^3P$) + CO$_2(j')$ elastic and inelastic quantum-mechanical scattering cross sections at collision energies between 0.03 and 5 eV. In addition, we present differential and momentum transfer cross sections, of interest for hot atomic oxygen transport and energy transfer at non-thermal-equilibrium conditions in middle and upper planetary atmospheres \citep{2019BAAS...51c.240K}.
Our newly constructed cross sections are of particular interest to Mars aeronomy missions, such as NASA's MAVEN mission \citep{2015SSRv..195....3J} and ESA's Trace Gas Orbiter (TGO), as well as to future Venus missions \citep{2011SSRv..162..267K}.
Moreover, velocity-dependent O($^3P$) + CO$_2(j)$ cross sections enter models seeking to explain variations in the mass-independent fractionation (MIF) of stratospheric carbon dioxide \citep{thiemens1995observation,wiegel2013unexpected} used to estimate carbon cycle reservoir size and rates of exchange for precise quantification of anthropogenic CO$_2$ and its effects on Earth's climate \citep{thiemens2014decadal}. 


\section{Theoretical Methods}
%
\subsection{Electronic Structure of \texorpdfstring{O($^3P$)-CO$_2$}{O(3P)-CO2} complex}

\begin{figure}
\centering
\noindent\includegraphics[width=0.6\columnwidth]{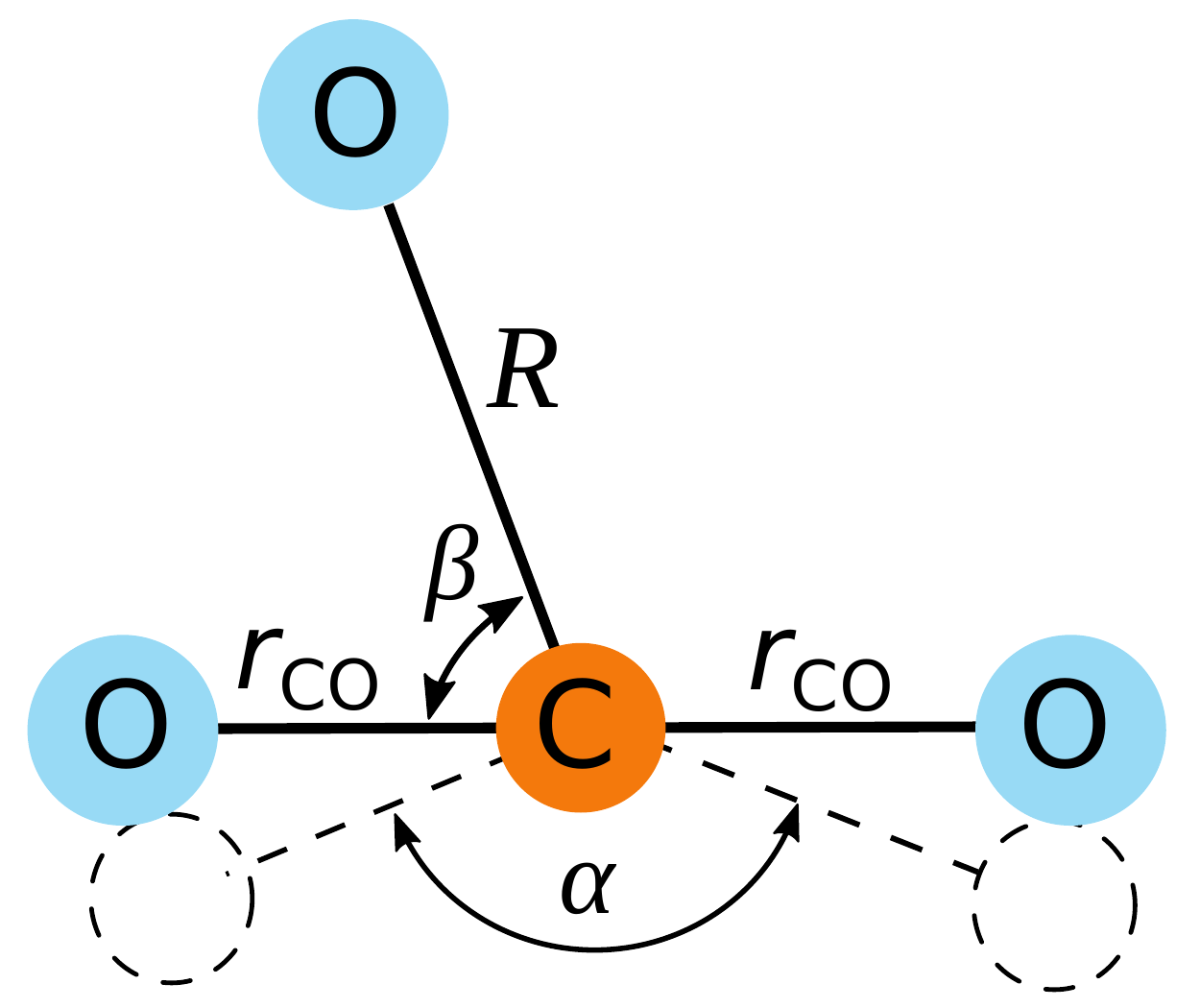}
\caption{Coordinate system used to parameterize the potential energy surfaces of the O($^3P$)+CO$_2$ complex. Planar symmetry is assumed. The bending angle $\alpha=180^\circ$ and $r_\mathrm{CO}=1.167$ $\AA{}$ at equilibrium distance for a non-vibrating CO$_2(00^{0}0)$ molecule.}
\label{fig:geometry}
\end{figure}

We first constructed \textit{ab-initio} potential energy surfaces (PESs) for three energetically lowest electronic states of the CO$_3$ complex that correlate to the O($^3P$) asymptote. 
We assumed that the CO$_{2}$ molecule is in its ground vibrational state, CO$_{2}(v_s v^{l}_b v_a)$=CO$_{2}(0 0^{0} 0)$, where $v_s=0$, $v^{l}_b=0$ and $v_a=0$ are symmetric, bending, and asymmetric vibrational modes, respectively. Non-vibrating CO$_{2}(0 0^{0} 0)$ is linear and symmetric, with the C-O internuclear distance, $r_{\mathrm{CO}}$, fixed at the equilibrium potential obtained by geometry optimization \citep{2006JChPh.124p4302D}.

\begin{figure*}
\noindent\includegraphics[width=1.0\textwidth]{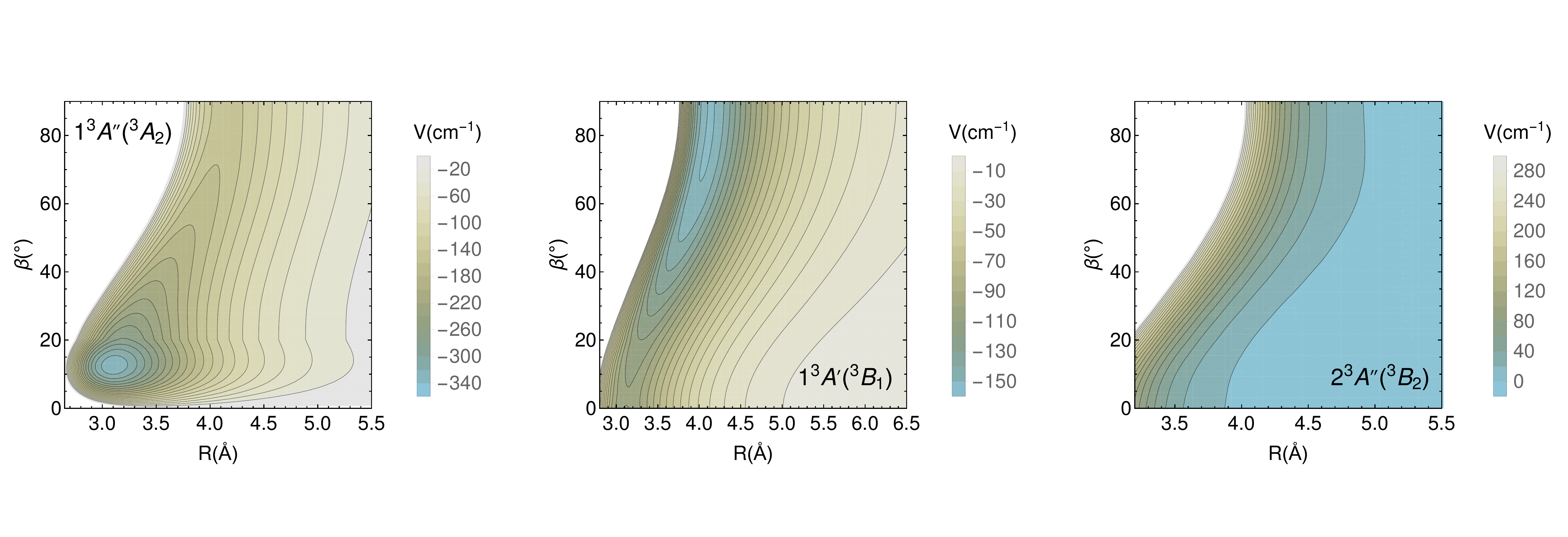}
\caption{Contour plots of the three energetically-lowest potential energy surfaces for the O($^3P$)-CO$_2$ complex: 1$^3A''$ (left), $^3A'$ (middle), 2$^3A''$ (right). The CO$_2$ molecule is kept at equilibrium in linear geometry. }
\label{fig:pes-contour}
\end{figure*}

With this assumption, the rotational symmetry with respect to the OCO internuclear axis is preserved and the three PESs of the CO$_3$ complex can be described in 2D planar geometry in terms of distance $R$, defined as the internuclear distance between the impacting O$(^3P)$ atom and the C atom of the CO$_2$ molecule, and the angle $0^\circ \leq \beta \leq 90^\circ$ between the CO$_2$ internuclear axis and the vector $R$ (Fig. \ref{fig:geometry}). The symmetry group of the system will be at least $C_S$ and depend on the angle of approach $\beta$. 
The lowest three PESs consist of a single $^3A'$ and two $^3A''$ electronic states. The special cases when the symmetry of the system is increased are the T-approach ($\beta=0^\circ$, symmetry group C$_{2v}$), and the colinear geometry ($\beta = 90^\circ$, symmetry group C$_{\infty v}$) (see \citet{2007MolPh.105.1171D} for details). 

We first evaluated the interaction energies of the complex at the spin-restricted single and double coupled-cluster with perturbative triple excitations (RCCSD(T)) level of theory \citep{2000JChPh.112.3106K} with the frozen core approximation on the augmented correlation-consistent polarized triple zeta (aug-cc-pVTZ-DK) atomic basis \citep{1989JChPh..90.1007D,1992JChPh..96.6796K}. 
We also carried out explicitly correlated calculations with partially spin-adapted scheme (RCCSD(T)-F12) \citep{knizia2009simplified} on a cc-pVTZ-F12 basis set \citep{2008JChPh.128h4102P}. Explicitly correlated calculations provide a dramatic improvement of the basis set convergence for coupled-cluster methods \citep{werner2010efficient,2011MolPh.109..407W}. 
Scalar relativistic effects for the system were observable and accounted for using the Douglas-Kroll-Hess Hamiltonian \citep{wolf2002generalized}.
To construct primitive surfaces, we kept the C-O internuclear distances frozen at their optimized equilibrium values ($r_\mathrm{CO}=1.167$ \AA{}) while varying the distance $R$, of the approaching O atom to the center of mass of the CO$_2$ molecule, on a non-equidistant grid with the highest point density around the global surface minima (if applicable). An equidistant grid in angle $\beta$ with steps of 10$^\circ$ was used. 
A total of about 200 \textit{ab-initio} points per surface were evaluated using both methods. All calculations were carried out using MOLPRO2012 \citep{MOLPRO,MOLPRO-WIREs}.

\begin{figure}
\noindent\includegraphics[width=\columnwidth]{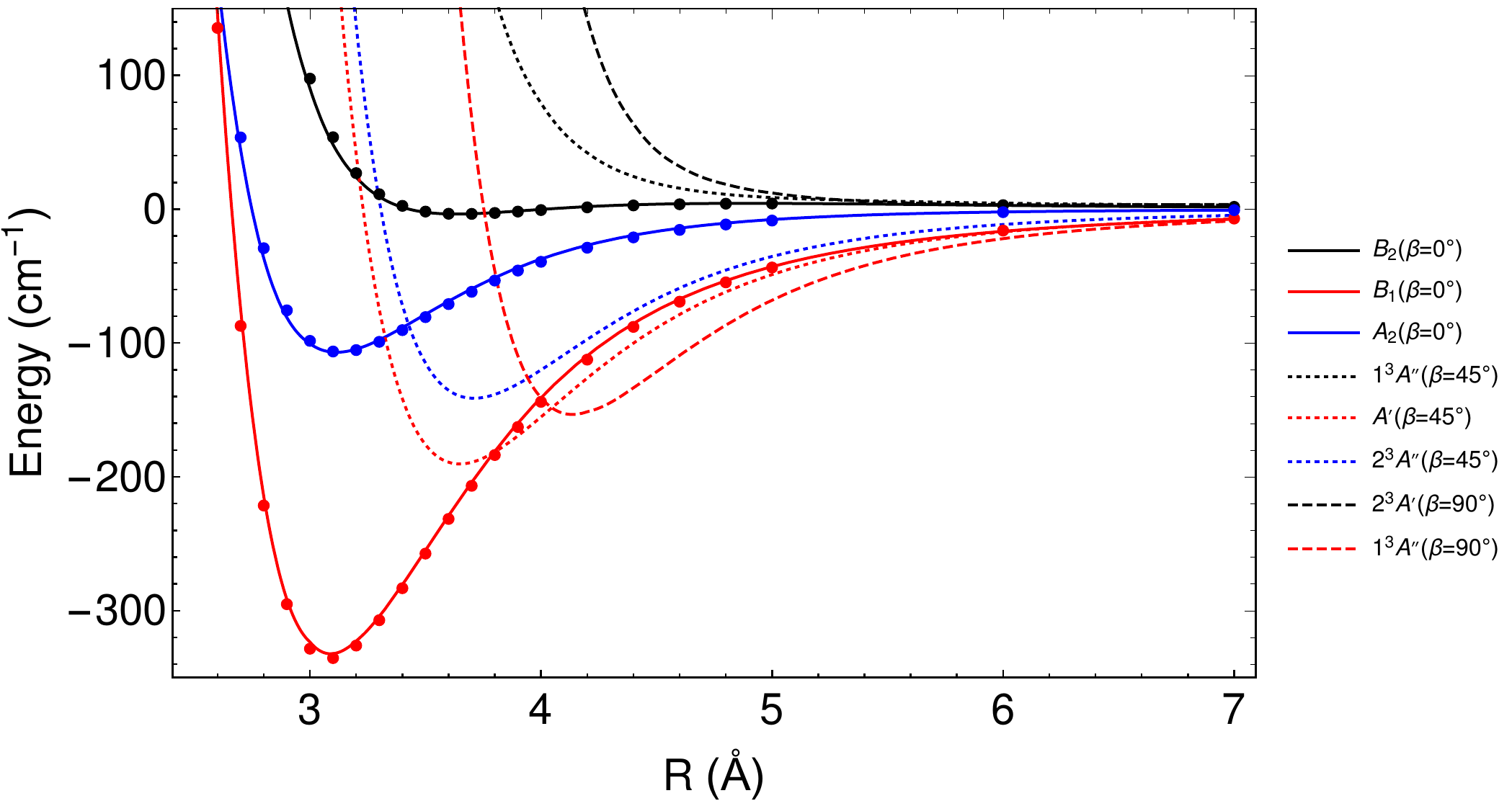}
\caption{One-dimensional plots of the constructed O($^3P$)-CO$_2$ potential energy surfaces for the selected angles: $\beta=0^\circ$ (T-approach; solid lines), $\beta=45^\circ$ (dotted lines), and $\beta=90^\circ$ (dashed lines). \textit{Ab-initio} points of \citet{2006JChPh.124p4302D} for $\beta=0^\circ$ are shown (filled circles) for comparison. CO$_2$ geometry is fixed as linear in equilibrium.}
\label{fig:pes-angles}
\end{figure}

Final PESs were generated from the \textit{ab-initio} points by constructing a fitting function of the form given by \citet{2012CPL...549...12S} first proposed by \citet{1999JChPh.110.3785B} (Fig. \ref{fig:pes-contour}).
We obtained excellent agreement with the PESs reported by \citet{2006JChPh.124p4302D} (Fig. \ref{fig:pes-angles}), with differences per energy point $<1$\% on average. We note that a common counterpoise technique \citep{boys1970calculation} to compensate for basis set superposition errors did not play a significant role, likely due to the large basis sets used and fast convergence of the explicitly correlated methods. 
A detailed report of our investigation of the electronic structure of CO$_3$ complex, including its dependence on the bending angles, falls out of the scope of the present work and will be published elsewhere.

\subsection{Collisional Dynamics}

We used MOLSCAT code \citep{MOLSCAT} to calculate state-to-state quantum mechanical elastic and inelastic cross sections for the scattering problem O($^3P$) + CO$_2(j) \rightarrow$ O($^3P$) + CO$_2(j')$, where the $j$ and $j'$ are the initial and final rotational quantum numbers of the CO$_2$ molecule, respectively.  
The resulting set of time-independent coupled-channel Schr\"odinger equations was solved in close-coupling formalism with CO$_2(j)$ molecule represented as a rigid rotor with electronic interaction O($^3P$) described by our newly constructed potential energy surfaces (given in previous section). The complete scattering $S$-matrices, and differential and integral cross sections were calculated separately for each of the three triplet PESs and statistically averaged to obtain the final values. 

We solved the scattering problem for 41 collision energy points on an equidistant grid between $E_\mathrm{col}=240$ cm$^{-1}$ [$2.976\times10^{-2}$ eV] and $E_\mathrm{col}=40240$ cm$^{-1}$ [4.9891 eV] in steps of 1000 cm$^{-1}$ [0.124 eV]. The reduced mass $\mu \equiv \mu_\mathrm{O,CO_2} = 11.73$ amu was used. 
Extensive testing was conducted to ensure the convergence with respect to the basis set size and numerical integration parameters. For the basis set consisting of rotational functions defined in terms of $j_\mathrm{max}$, the maximum allowed rotational quantum number of CO$_2$ molecule, we found that satisfactory convergence of the elastic cross sections can be achieved with a basis set as small as $j_\mathrm{max}=30$. This remained true even when the centrifugal sudden (CS) approximation \citep{mcguire1974quantum} was invoked to reduce the calculation size by approximating the transitions between rotational states higher than $\Delta j=2$ (JZCSMX=2). 
Nevertheless, in order to ensure convergence of inelastic cross sections for initial rotational states $j$, present in the Maxwell-Boltzmann tail for the thermal upper atmosphere of Mars, we carried out production runs using a much larger basis set, with $j_\mathrm{max}=100$ and JZCSMX=8.
The convergence with respect to $J$, the total rotational quantum number of the complex, was achieved for $J_\mathrm{max} = 200-900$, where $J_\mathrm{max}$ increases with collision energy.

\begin{table}
\caption{Numerical integration parameters used in MOLSCAT production runs.}
\centering
\begin{tabular}{l r}
\hline
 Parameter          & Value  \\
\hline
  $j_\mathrm{max}$  & 100          \\
  $J_\mathrm{max}$  & 900          \\
  JZCSMX            & 8            \\
  RMIN (\AA{})      & 0.7          \\
  RMAX (\AA{})      & 30$^a$        \\
  RVFAC             & 1.3           \\
  STEPS             & 12            \\
\hline
\multicolumn{2}{l}{$^{a}$Automatic scaling with $J$ was used.}
\end{tabular}
\label{table_molscat}
\end{table}

Numerical integration was performed using modified log-derivative \citep{1986JChPh..85.6425M} and hybrid modified log-derivative/Airy (LDA) \citep{alexander1987stable} propagators. We found that satisfactory precision could be achieved using the faster LDA scheme, with Airy propagator set up to automatically detect integration region boundaries. The production run parameters are given in Table \ref{table_molscat}. 

\section{Results and Analysis}

\subsection{Elastic and inelastic cross sections}

\begin{figure*}
  \centering
  \noindent\includegraphics[width=0.7\textwidth]{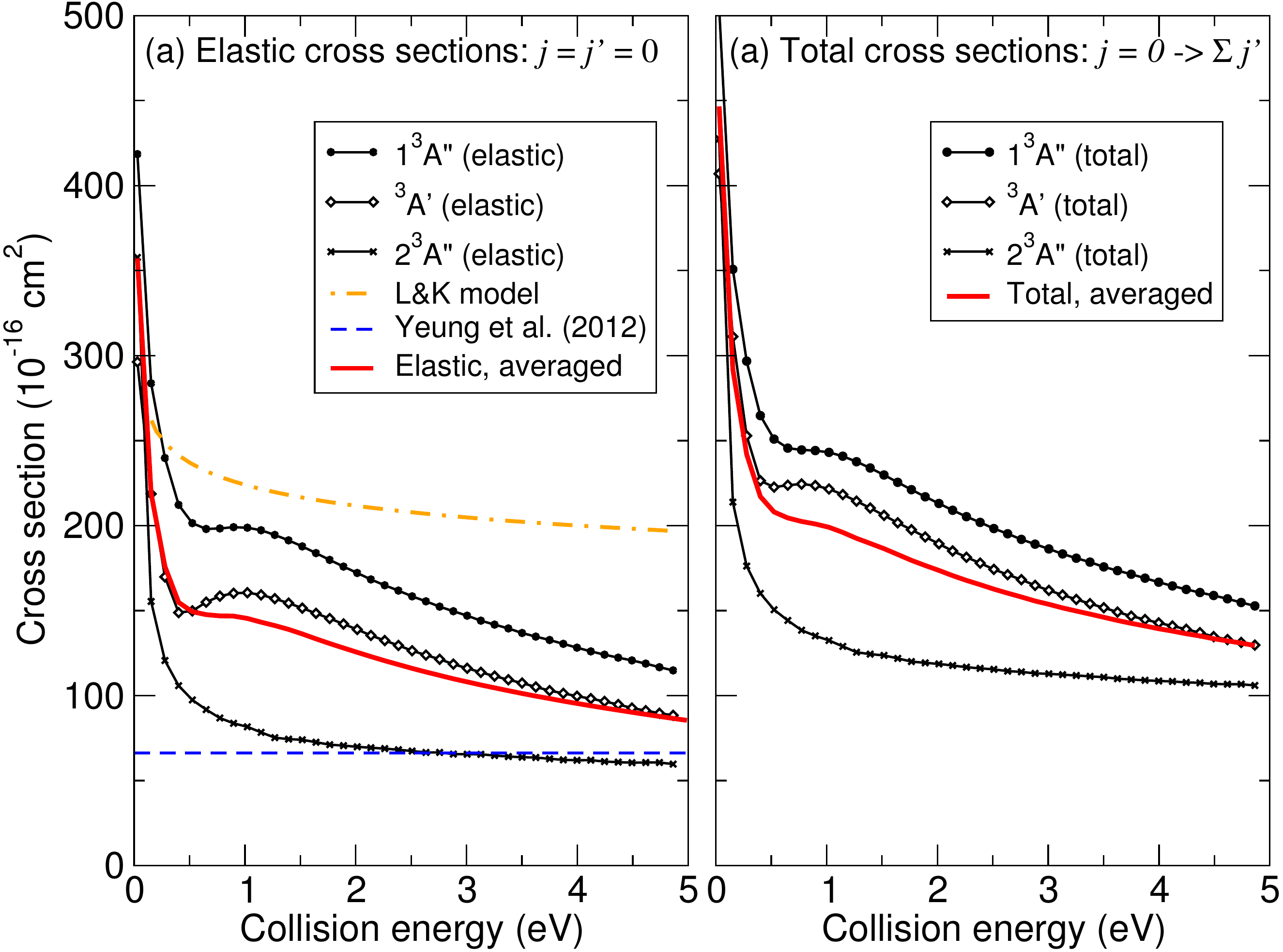}
  \caption{Cross sections for O($^3P$) + CO$_2$($j$=0) $\rightarrow$ O($^3P$) + CO$_2(j')$ scattering on three PESs: 1$^3A''$ (circles), $^3A'$ (diamonds), 2$^3A''$ (crosses), and their statistically weighted average (solid line). \textit{Left:} Elastic cross sections for $j$=0 compared to L\&K model \citep{2014ApJ...790...98L} (dot-dashed) and hard-sphere cross section \citep{2012JPCA..116...64Y} (dashed). \textit{Right:} Total (elastic + inelastic) cross sections $\sigma_{j=0}(E)$ = $\sum_{j'} \sigma_{j'}(E)$, where $j'=0$ to $j_\mathrm{max}$.}
  \label{fig:elastic-total-cs}
\end{figure*}

For the three PESs, we calculated elastic and inelastic state-to-state cross sections, $\sigma_{j,j'}(E)$, for the O($^3P$) + CO$_2(j) \rightarrow$ O($^3P$) + CO$_2(j')$ scattering, where rotational quantum numbers $j$ and $j'$ range from 0 to $j_\mathrm{max}$. The total scattering cross sections on a surface is given by  
\begin{equation}
  \sigma_{j}(E) = \sum_{j'} \sigma_{j,j'}(E) , 
\end{equation} 
where $j'=0 \ldots j_\mathrm{max}$. The cross sections are shown in Fig. \ref{fig:elastic-total-cs} as a function of the collision energy, $E=0.03-5$ eV in the center-of-mass frame of reference, as well as in Table \ref{table_cs1} for selected collision energies and rotational states.

\begin{table*}
\caption{State-to-state cross sections $\sigma_{j=0,j'}(E)$ (units of $10^{-16}$ cm$^2$) for selected energies.}
\centering
\begin{tabular}{r|rrrr rrrrr}
\hline
        & \multicolumn{9}{c}{Center-of-mass collision energy $E$ (eV)} \\
 $j'$   & 0.0297 & 0.154 & 0.28 & 0.53  & 0.77 & 1.51 & 2.5 & 3.5 & 4.0 \\
\hline
0	& 357.16	& 218.851	& 175.952	& 146.854	& 145.377	& 136.222	& 116.021	& 101.273	& 95.37 \\
2	& 40.858	& 25.328	& 20.963	& 15.329	& 13.266 & 10.877	& 8.609		& 7.512		& 7.122 \\
4	& 14.632	& 9.308		& 7.779		& 4.616		& 3.974	& 3.294	& 2.78	& 2.489		& 2.426 \\
6	& 9.495	& 6.425	& 5.182	& 3.433	& 3.084	& 2.686	& 2.484	& 2.389	& 2.367 \\
8	& 6.674	& 4.799	& 3.94	& 2.663	& 2.426	& 2.083	& 1.838	& 1.801	& 1.815 \\
10	& 4.23	& 3.094	& 2.417	& 1.862	& 1.713	& 1.628	& 1.457	& 1.369	& 1.367 \\
12	& 2.918	& 2.701	& 2.497	& 2.093	& 1.942	& 1.74	& 1.586	& 1.477	& 1.452 \\
14	& 2.91	& 2.61	& 2.337	& 1.798	& 1.683	& 1.413	& 1.304	& 1.275	& 1.265 \\
16	& 2.11	& 2.101	& 1.739	& 1.36	& 1.258	& 1.173	& 1.056	& 1.017	& 0.983 \\
18	& 1.96	& 1.988	& 1.799	& 1.521	& 1.415	& 1.406	& 1.311	& 1.18	& 1.142 \\
20	& 1.721	& 1.838	& 1.645	& 1.291	& 1.37	& 1.289	& 1.051	& 0.994	& 0.941 \\
22	& 0.901	& 1.581	& 1.354	& 1.052	& 1.128	& 1.014	& 0.933	& 0.87	& 0.868 \\
24	& 0.31	& 1.564	& 1.443	& 1.252	& 1.16	& 1.21	& 1.049	& 0.984	& 0.947 \\
26	& 	& 1.42	& 1.291	& 1.057	& 1.049	& 1.103	& 0.875	& 0.875	& 0.888 \\
28	& 	& 1.051	& 1.009	& 0.848	& 0.974	& 0.92	& 0.761	& 0.74	& 0.716 \\
30	& 	& 1.192	& 1.15	& 0.976	& 1.027	& 0.978	& 0.891	& 0.846	& 0.837 \\
32	& 	& 1.1	& 1.076	& 0.853	& 0.822	& 0.878	& 0.785	& 0.736	& 0.712 \\
34	& 	& 0.823	& 0.759	& 0.679	& 0.739	& 0.822	& 0.69	& 0.707	& 0.71 \\
36	& 	& 0.647	& 0.78	& 0.788	& 0.918	& 0.852	& 0.818	& 0.812	& 0.772 \\
38	& 	& 0.661	& 0.83	& 0.75	& 0.786	& 0.781	& 0.7	& 0.646	& 0.667 \\
40	& 	& 0.636	& 0.702	& 0.594	& 0.613	& 0.72	& 0.668	& 0.627	& 0.618 \\
42	& 	& 0.497	& 0.578	& 0.642	& 0.728	& 0.73	& 0.749	& 0.728	& 0.737 \\
44	& 	& 0.408	& 0.608	& 0.668	& 0.702	& 0.684	& 0.793	& 0.692	& 0.701 \\
46	& 	& 0.376	& 0.572	& 0.563	& 0.541	& 0.596	& 0.633	& 0.583	& 0.584 \\
48	& 	& 0.282	& 0.474	& 0.53	& 0.557	& 0.653	& 0.687	& 0.681	& 0.688 \\
50	& 	& 0.191	& 0.425	& 0.56	& 0.595	& 0.62	& 0.662	& 0.609	& 0.579 \\
52	& 	& 0.148	& 0.443	& 0.498	& 0.511	& 0.504	& 0.594	& 0.579	& 0.59 \\
54	& 	& 0.111	& 0.396	& 0.452	& 0.494	& 0.575	& 0.65	& 0.608	& 0.554 \\
56	& 	& 	& 0.331	& 0.462	& 0.496	& 0.546	& 0.619	& 0.603	& 0.589 \\
58	& 	& 	& 0.299	& 0.448	& 0.465	& 0.472	& 0.557	& 0.575	& 0.537 \\
60	& 	& 	& 0.295	& 0.393	& 0.432	& 0.468	& 0.557	& 0.591	& 0.556 \\
62	& 	& 	& 0.247	& 0.387	& 0.447	& 0.496	& 0.569	& 0.579	& 0.583 \\
64	& 	& 	& 0.187	& 0.394	& 0.411	& 0.466	& 0.474	& 0.522	& 0.535 \\
66	& 	& 	& 0.138	& 0.372	& 0.39	& 0.413	& 0.466	& 0.528	& 0.549 \\
68	& 	& 	& 0.112	& 0.338	& 0.385	& 0.434	& 0.48	& 0.515	& 0.532 \\
70	& 	& 	& 0.077	& 0.338	& 0.376	& 0.413	& 0.498	& 0.537	& 0.517 \\
72	& 	& 	& 0.036	& 0.34	& 0.356	& 0.376	& 0.444	& 0.468	& 0.506 \\
74	& 	& 	& 0.008	& 0.296	& 0.334	& 0.398	& 0.456	& 0.456	& 0.456 \\
76	& 	& 	& 	& 0.264	& 0.317	& 0.418	& 0.429	& 0.447	& 0.45 \\
78	& 	& 	& 	& 0.282	& 0.324	& 0.366	& 0.405	& 0.424	& 0.427 \\
80	& 	& 	& 	& 0.275	& 0.332	& 0.348	& 0.407	& 0.442	& 0.471 \\
82	& 	& 	& 	& 0.263	& 0.293	& 0.368	& 0.416	& 0.476	& 0.466 \\
84	& 	& 	& 	& 0.248	& 0.273	& 0.343	& 0.392	& 0.394	& 0.389 \\
86	& 	& 	& 	& 0.224	& 0.285	& 0.31	& 0.411	& 0.426	& 0.44 \\
88	& 	& 	& 	& 0.235	& 0.276	& 0.326	& 0.422	& 0.427	& 0.447 \\
90	& 	& 	& 	& 0.216	& 0.251	& 0.356	& 0.359	& 0.421	& 0.476 \\
92	& 	& 	& 	& 0.208	& 0.268	& 0.289	& 0.372	& 0.47	& 0.448 \\
94	& 	& 	& 	& 0.225	& 0.249	& 0.308	& 0.447	& 0.434	& 0.432 \\
96	& 	& 	& 	& 0.195	& 0.272	& 0.359	& 0.346	& 0.357	& 0.359 \\
98	& 	& 	& 	& 0.221	& 0.269	& 0.253	& 0.328	& 0.413	& 0.44 \\
100$^a$	& 	& 	& 	& 0.094	& 0.215	& 0.391	& 0.569	& 0.669	& 0.685 \\
\hline
\multicolumn{10}{l}{$^{a}$Contributions from all $j'>100$.}
\end{tabular}
\label{table_cs1}
\end{table*}

With the exception of the lowest energy point ($E=0.0297$ eV), our elastic cross sections, $\sigma_{0,0'}(E)$, are greater than estimated scattering cross sections of \citet{2012JPCA..116...64Y}, who based their value on a Lennard-Jones interaction potential constructed on the lowest surface with parameters estimated from the experiment, but smaller than the elastic cross sections of \citet{2014ApJ...790...98L} and semi-empirical values by \citet{2014Icar..228..375F,2018Icar..300..411F} used in numerous recent studies of Mars aeronomy. 
Specifically, at $E=0.5$ eV and $E>4$ eV, our elastic cross sections appear to be smaller by more than 25\% and 50\%, respectively, than reported in the above studies. 
Moreover, our total cross sections are about 25\% larger than the elastic cross sections, suggesting that internal (rotational) degrees of freedom of CO$_2$ molecule are significantly populated in collisions.

\begin{figure}
  \centering
  \noindent\includegraphics[width=\columnwidth]{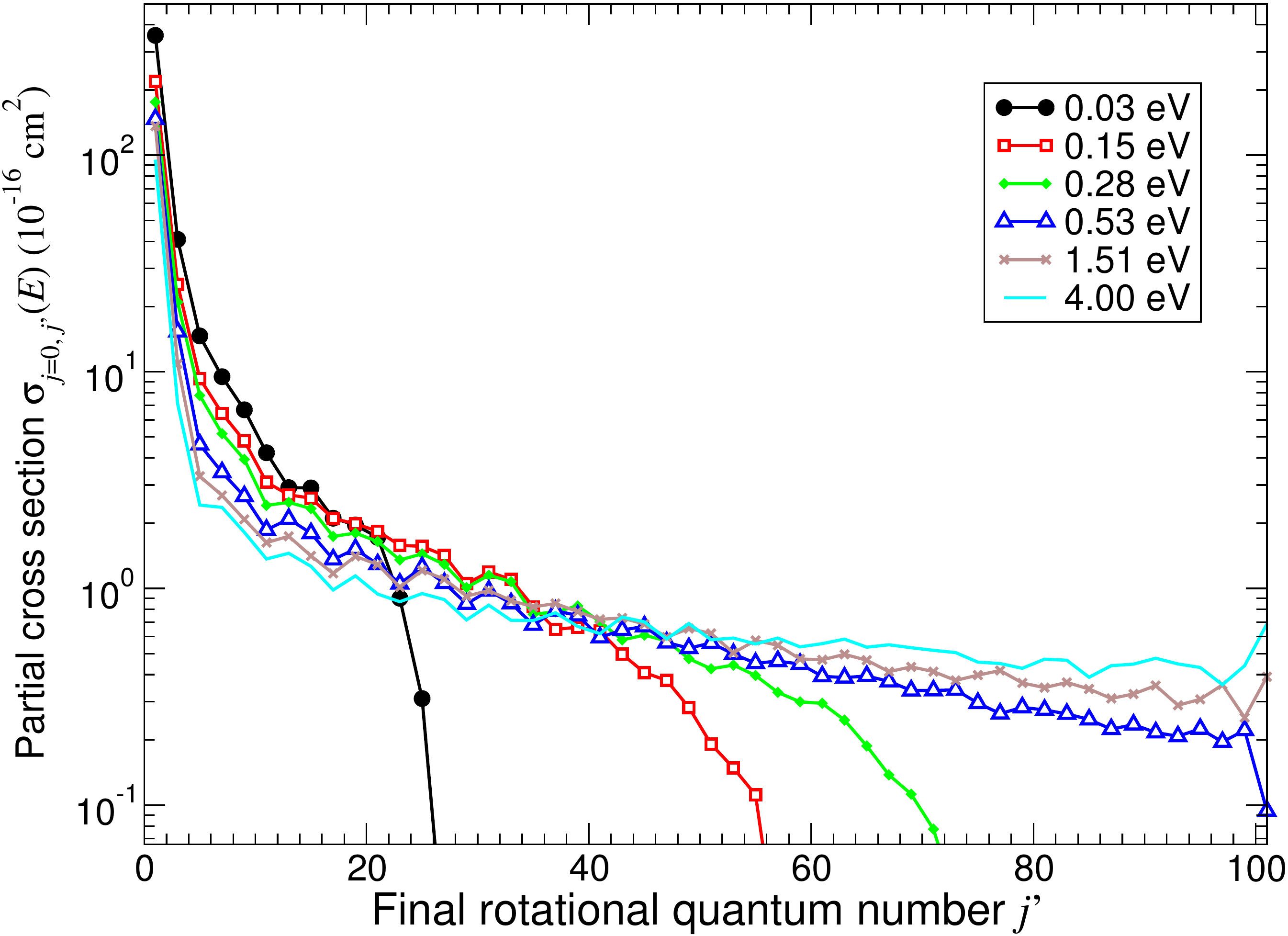}
  \caption{Cross sections $\sigma_{j=0,j'}(E)$ for O($^3P$) + CO$_2$(j=0) $\rightarrow$ O($^3P$) + CO$_2(j')$ scattering on statistically averaged PES given as a function of $j'$ for selected collision energies. Estimated uncertainties due to finite basis size are shown as residual values at $j'=101$ (last point).}
  \label{fig:inelastic1}
\end{figure}

The relative significance of inelastic excitations is illustrated in Fig. \ref{fig:inelastic1}, where we show cross sections $\sigma_{j=0,j'}(E)$ as a function of rotational quantum number $j'$ for selected collision energies. 
Low-energy O atoms are 2-10 times more efficient at exciting low rotational states of CO$_2$, but do not carry enough kinetic energy to excite high rotational states. This is consistent with general predictions of scattering theory where longer interaction times lead to more efficient energy transfer. 

At collision energies smaller than about 0.5 eV, the rotational basis set is sufficiently large to fully capture the scattering dynamics (\textit{i.e.}, the collision energies do not exceed the highest rotational energy included in the basis set), resulting in the cutoff in the rotational excitation spectrum. For example, at collision energy $E=0.28$ eV, rotational states up $j' \approx 78$ can be excited (Fig. \ref{fig:inelastic1}). 

For collision energies higher than about 0.5 eV, all rotational channels are collisionally populated and the approximate contributions from $j'>j_\mathrm{max}$ are included in $\sigma_{j,j_\mathrm{max}}(E)$. 
This approach allows us to estimate the uncertainties introduced by the finite basis size for $j_\mathrm{max}=100$ used in the production runs to be smaller than 0.1\%, 
and largely negligible for low rotational levels (\textit{e.g.}, $j'=0-5$). For high rotational levels (\textit{e.g.}, $j'>15-20$), the uncertainty increases with collision energy and can be estimated to about 1-2\% for $j'>20-40$ and up to 10\% at $j'=90-100$ at $E>4$ eV.

\subsection{Differential cross sections}

\begin{figure*}
  \centering
  \noindent\includegraphics[width=0.9\textwidth]{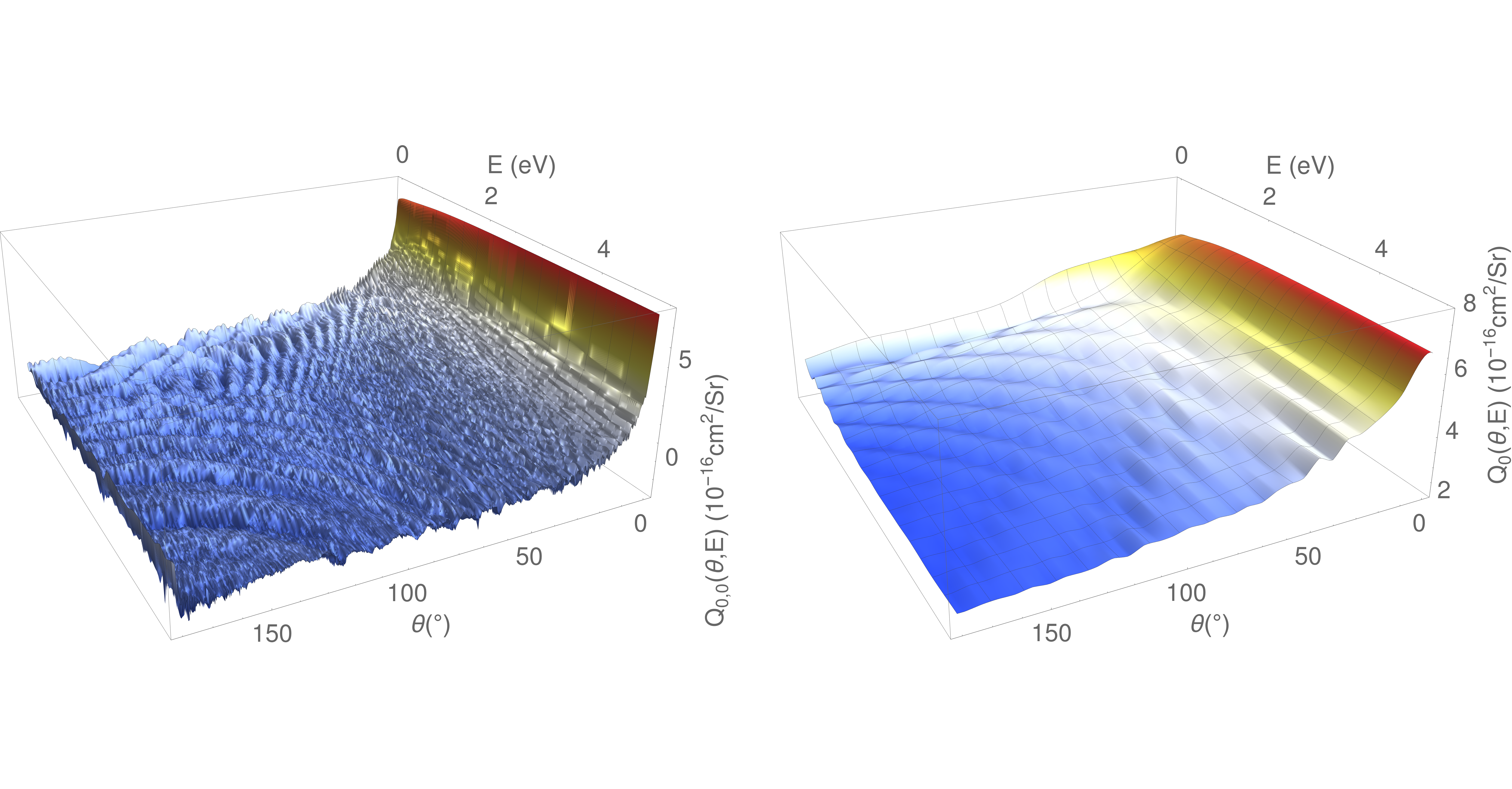}
  \caption{Elastic (left panel) and total (right panel) differential cross sections for O($^3P$) + CO$_2(j = 0)$ scattering, statistically averaged over all PESs and presented in log scale.}
  \label{fig:DCS}
\end{figure*}

\begin{figure}
  \centering
  \noindent\includegraphics[width=\columnwidth]{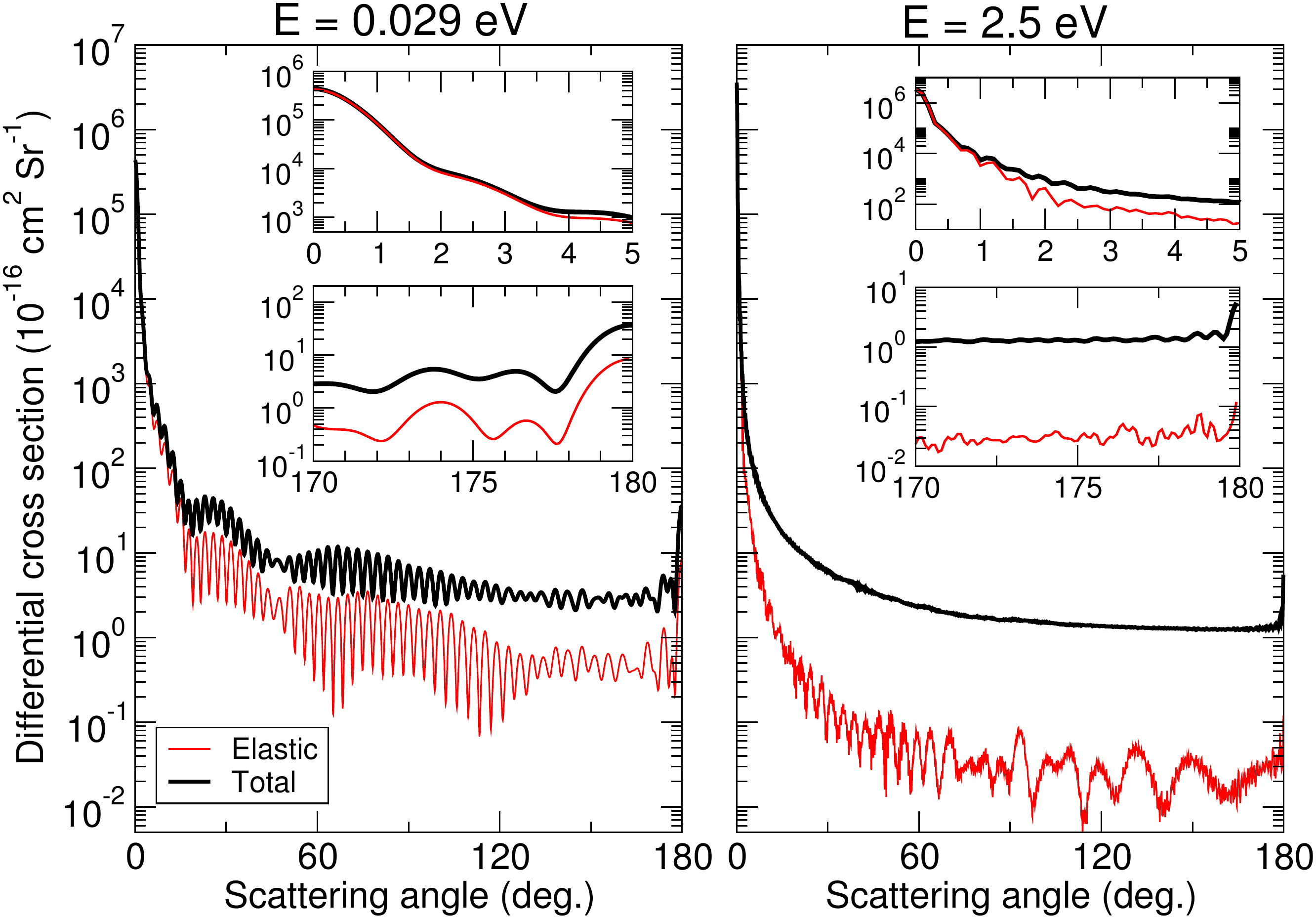}
  \caption{Elastic (thin red line) and total (thick black line) DCSs as functions of scattering angle shown for collision energies $E$=0.0297 eV (left panel) and $E$=2.5 eV (right panel). \textit{Insets:} Zoom in on small (top) and large (bottom) scattering angles.}
  \label{fig:DCS_selectedE}
\end{figure}

State-to-state differential cross sections (DCSs) were calculated from the numerically evaluated scattering $S$-matrices over the considered collision energy range using the MOLSCAT utility code \texttt{dcssave.f}. The DCSs were constructed independently for each of the three PESs.
The elastic DCSs, $Q_{j,j'}(\theta,E) = {d \sigma_{j,j'}(\theta,E)}/{d \Omega}$ and total DCSs, $Q_{j}(\theta,E) = \sum_{j'}{d \sigma_{j,j'}(\theta,E)}/{d \Omega}$, are illustrated in Fig. \ref{fig:DCS} for $j=0$ in dependence on the center-of-mass collision energy $E$ and the scattering angle $\theta$. Here, $d\Omega = \sin \theta d \theta d \phi$ is the solid angle element and $\theta$ is the scattering angle, defined as the relative angle between the incoming and outgoing trajectories of the incoming O($^3P$) atom. In other words, $\theta=0^\circ$ corresponds to complete forward scattering without deflection, $\theta=90^\circ$ to sideways scattering, and $\theta=180^\circ$ to complete backscattering. 
Example DCSs for two collision energies, 0.0297 eV and 2.5 eV, representative of low- and high-energy scattering, are given in Fig. \ref{fig:DCS_selectedE}

Both elastic and total DCSs are strongly anisotropic, with a dominant forward-scattering peak ($0^\circ < \theta < 2^\circ$ contributes to more than 90\% of the integral cross section) and backscattering peaks clearly visible. Note that integral cross sections were calculated for the entire range of the scattering angle, $0^\circ < \theta < 180^\circ$.
For small scattering angles $\theta$, the DCSs are almost purely elastic. The inelastic excitations start to appear for $\theta>3^\circ$ for low-energy collisions and for $\theta>1^\circ$ for high-energy collisions (Fig. \ref{fig:DCS_selectedE}, inset). 
For larger scattering angles the total DCSs are up to two orders of magnitude greater than the elastic DCSs, suggesting that rotational excitations are dominant, in particular at higher collision energies.
The oscillatory structure present in elastic cross sections, in particular at low collision energies, is a real feature that does not average out because the number of participating partial waves is much smaller than in case of total DCSs, where both elastic and inelastic channels are included. At higher collision energies the structure is smoothed out due to a much larger number of contributing partial waves. Note that the values shown in Fig. \ref{fig:inelastic1} are for a single energy rather than averaged out over a realistic velocity distribution. 
At low collision energies several wave envelopes with maxima at scattering angles $\theta$ equal to about $30^\circ$, 70$^\circ$, 100$^\circ$, and 140$^\circ$ can be resolved.  
The complete dataset containing state-to-state CSs and DCSs is available online  \citep{github_repo} as well as on request.

\subsection{Momentum transfer cross section}

Momentum transfer (or momentum \textit{transport}) cross section (MTCS) is an effective quantity suitable for modeling the average momentum transferred from a projectile to the target particle in a binary collision. MTCSs are used in studies of energy exchange, diffusion, and transport in non-thermal environments in astrophysics, atmospheric science, and aeronomy \citep{1997JASTP..59..107K,2009JGRA..114.7101Z}.

\begin{figure}
  \centering
  \noindent\includegraphics[width=\columnwidth]{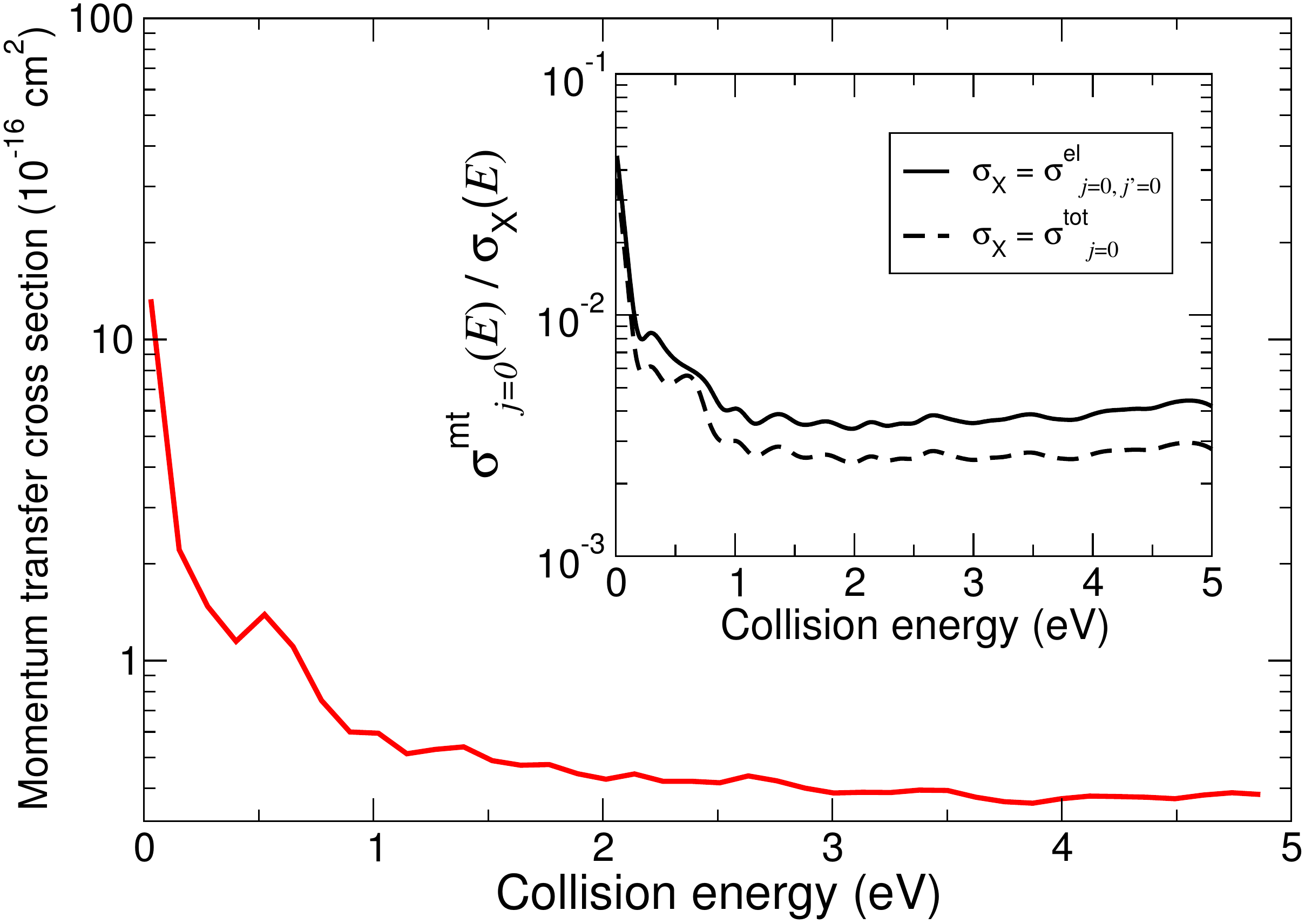}
  \caption{Momentum transfer cross section $\sigma_{j=0}^{\mathrm{mt}}(E)$ for O($^3P$) + CO$_2$($j$ = 0) as a function of collision energy statistically averaged over three PESs. 
  \textit{Inset:} Ratio of the MTCS and elastic and total cross sections, respectively, for the same initial state. }
  \label{fig:mtcs}
\end{figure}

For O($^3P$) + CO$_2(j = 0) \rightarrow$ O($^3P$) + CO$_2(j'=0)$ elastic scattering, the momentum transfer cross section is given by
\begin{equation}
 \sigma_{j=0}^{\mathrm{mt}}(E) = \int_0^\infty Q_{0,0}(\theta,E) \sin \theta (1-\cos \theta) \mathrm{d} \theta  \; ,
 \label{eq:mtcs}
\end{equation}
where $Q_{0,0}(\theta,E)$ is the differential cross section for $j=j'=0$.
Following Eq. (\ref{eq:mtcs}), we have calculated MTCSs as a function of collision energy independently for each PES, as well as their statistically weighted average (Fig. \ref{fig:mtcs}).

Computed MTCSs are on average less than 1\% of the elasic and total cross sections, except for low energy collisions, indicating that energy transfer from superthermal O to thermal CO$_2$ background is rather inefficient (Fig. \ref{fig:mtcs}, \textit{inset}) even in comparison with O($^3P$)+H$_2$ collisions where MTCSs are greater than 2\% \citep{2012GeoRL..3910203G}. In comparison, in the same collision energy range, MTCSs for atom-atom collisions tend to be about 10\%-25\% of the total cross sections between a heavy projectile and a light target species such as N+He, and about 2\%-18\% for heavy species, such as N+Ar \citep{PhysRevLett.100.103001}.
The evaluated MTCSs have direct consequences on Mars aeronomy and atmospheric escape, where we can expect larger escape fluxes of non-thermal atomic oxygen produced at lower altitudes \citep{2017JGRA..122.3815L}, and possibly other atomic species, such as C and N, escaping through a thick layer of CO$_2$.

%

\section{Summary and Discussion}

We have carried out a theoretical study of O($^3P$) + CO$_2 \rightarrow$ O($^3P$) + CO$_2$ non-reactive scattering at thermal and superthermal collision energies up to 5 eV in the center-of-mass frame. The selected energy range is typical in non-thermal-equilibrium conditions in upper planetary atmospheres exposed to solar radiation and plasma inputs.
To describe the electronic interactions between the colliding particles, we have constructed \textit{ab-initio} potential energy surfaces for three electronic states correlating to the energetically-lowest asymptote of the O($^3P$)-CO$_2(^1\Sigma^+)$ complex. The surfaces were constructed in a restricted planar geometry, with the CO$_2$ molecule assumed to be linear with C-O internuclear separation fixed at the equilibrium distance.

With the PESs as inputs, we computed velocity-dependent state-to-state and total elastic and inelastic cross sections, corresponding differential cross sections, as well as elastic momentum transfer cross sections. 
The CO$_2$ molecule was modeled as a rigid rotor and rotational-vibrational couplings between the symmetric, bending, and asymmetric vibrational modes were not included in the production run of the model. The scattering problem was described using a quantum mechanical coupled-channel formalism and solved numerically. Thus, the cross sections and derived physical quantities were constructed from first principles without any external empirical parameters. Although computationally intensive, the \textit{ab initio} approach has several major advantages over classical theory, such as evaluation of all state-to-state transitions induced by collisions and more complete treatment of purely quantum effects, expected to play a significant role in improving the predictive power and precisions of models \citep{2019BAAS...51c.240K}. 

Our computed elastic cross sections are smaller than the elastic cross sections estimated by mass-scaling from well-described atom-molecule systems \citep{2014ApJ...790...98L,2018Icar..300..411F}, but larger than the Lennard-Jones cross section reported by \citet{2012JPCA..116...64Y}. 
Specifically, at superthermal energies, our elastic cross sections range between $1.5\times10^{-14}$ cm$^2$ at about 1 eV collision energy, down to $8.5\times10^{-15}$ cm$^2$ at 5 eV. In comparison, the widely used elastic cross sections of \citet{2014ApJ...790...98L} evaluate to $2.24\times10^{-14}$ cm$^2$ and $1.97\times10^{-14}$ cm$^2$  for the two collision energies, respectively. 
Calculated differential cross sections (DCSs) are strongly favoring forward scattering with the anisotropy increasing with the collision energy.
The small-angle forward scattering is mostly elastic with negligible kinetic energy transfer to internal degrees of freedom of the CO$_2$ molecule. 
In contrast, for larger scattering angles the inelastic DCSs are nearly two orders of magnitude greater than elastic ones, indicating that the kinetic energy transfer to internal excitations is rather efficient. For the highest collision energies considered, the total inelastic CS is equal to approximately one half of the elastic CS, suggesting that up to 50\% of the translational kinetic energy of the O($^3P$) atom is converted into internal excitations (rotations) of the CO$_2$ molecule. 
\citet{2012JPCA..116...64Y} observed significant energy transfer in the O$(^3P)$ + CO$_2$ system, with ``an average of only 41\% of the collision energy remaining in product translation''. A simple statistical estimate based on level energies puts that number as high as 75\% \citep{2018Icar..300..411F}. 
In comparison, in O($^3P$)+CO collisions, on average 84\% of the energy remains in product translation, approximately 80\% is retained in Ar + ethane collisions, and 60\% in superthermal O($^3P$)+C$_2$H$_6$ collisions.
\citet{2012JPCA..116...64Y} suggested that such a large fraction of kinetic energy transfer to inelastic excitations may be due to significant contribution from a short-lived intermediate CO$_3$ complex that can form in the collisions and yield reaction products corresponding to inelastic scattering of O$(^3P)$+CO$_2$. 

The most significant contribution to the uncertainties in the calculated cross sections comes from approximating the CO$_2$ molecule with a linear rigid rotor. With vibrational modes and rotational-vibrational coupling included in the model, we expect our inelastic (and total) scattering cross sections to be lower than what nature intended because of absence of direct kinetic energy transfer into vibrational modes. The effects should be more significant at the high-end of the collision energy range considered, above about 4 eV, where the vibrational excitations were predicted to overtake rotational excitations as a  dominant inelastic process \citep{schatz1981quasiclassical}.

Based on energy transfer arguments and reported vibrational cross sections in existing studies, we can attempt to estimate the uncertainties of our results. If we require 65\% of translational energy to be converted into internal excitations of CO$_2$ molecule at collision energy $E=4.3$ eV, as proposed by \citet{2012JPCA..116...64Y}, our total cross sections should be increased by about 13\%, to $\sigma_{j=0} \approx 1.5\times10^{-14}$ cm$^2$. Similarly, at lower collision energies, our total cross sections should be increased by a proportionally smaller percentage, as per energy arguments. Note that the estimated fraction of the translational energy transferred to the internal excitations by \citet{2012JPCA..116...64Y} is rather high.

In comparison, a theoretical study of vibrational excitations in the title process based on the infinite-order-sudden (IOS) approximation \citep{1982CPL....88..553M} reported cross sections for collisionally exciting CO$_2(01^10)$ (first bending mode) greater than $5.6 \times 10^{-17}$ cm$^2$ at $E>2$ eV, or less than 1\% of our elastic cross sections.
\citet{1992JChPh..96.2025U} measured the cross sections for O-atom excitation of CO$_2(00^01)$ mode at $E=3.9$ eV (8 km/s) to be equal to $3.7 \times 10^{-18}$ cm$^2$ and recommended a value of $6.4\times 10^{-17}$ cm$^2$ as a cross section for O-atom excitation of CO$_2(01^10)$. Their proposed values are in good agreement with a quasi-classical trajectory (QCT) study of collisional excitation of CO$_2$($NN'1$), where $N$ and $N'$ correspond to excited symmetric and bending vibrational modes, respectively \citep{schatz1981quasiclassical}.
These studies suggest that the cross sections for collisional excitation of the lowest antisymmetric vibrational mode of CO$_2$ are, at best, nearly two orders of magnitude smaller than our inelastic (rotational) cross sections. 
If we assume that the total inelastic cross section for vibrational excitations scales linearly with the number of open channels (vibrational states), we can make a more direct comparison with our results. At collision energy $E=3.9$ eV, lowest 14 antisymmetric vibrational modes, up to $v_a=13$, can be excited, yielding a total vibrational cross section $8.3\times10^{-16}$ cm$^2$ for CO$_2(N N',13)$, or about one half of the value estimated from the translational energy fractions.

The elastic cross sections are less likely to be significantly affected by the absence of vibrational excitations in our model. We tested the convergence of scattering calculations for basis sets comprised of as few as six internal states $(j'_\mathrm{max}=5)$, or of only 5-10 bending mode energies ($E_{b} = 667$ cm$^{-1}$) with rotational states modeled using the IOS approximation \citep{goldflam1977infinite,1979JChPh..70.4686G}. The elastic cross sections computed with these minimal basis sets were about 20\% larger than our production values, while a basis set with as few as 30 rotational states matched the production values within 2\%. Nevertheless, matching the production inelastic cross sections required large basis sets.

Three final points should be mentioned. First, deviation from the linear geometry (bending angle $\alpha \neq 180^\circ$) will likely have an effect on the cross sections. \citet{2006JChPh.124p4302D,2007MolPh.105.1171D} constructed PESs for the bending angles $150^\circ < \alpha < 210^\circ$ and demonstrated that for $\alpha < 180^\circ$, for which the CO$_2$ molecule bends towards the incoming O($^3P)$ atom, the surfaces become more attractive and have deeper wells, while the opposite is true for $\alpha > 180^\circ$. However, since the binding angle is directly connected to the strength of rotational-vibrational couplings, it is difficult to assess the effects of geometry changes without conducting further studies with vibrational degrees of freedom included in the model. Our preliminary calculations match well the PESs of \citet{2006JChPh.124p4302D} for bending angles $150^\circ < \alpha < 210^\circ$ and the work to extend the scattering model to include vibrations is in progress. 
Second, even though we conducted the scattering calculations on three PESs correlating to the lowest energy asymptote of O$(^3P)$ + CO$_2$ complex, we did not take into account intersurface couplings that could lead to surface hopping. Such studies were conducted for spin-orbit couplings at lower collision energies \citep{2007MolPh.105.1171D} and the impact of surface hopping on cross sections at similar scattering energies were analyzed for other atom-molecule systems \citep{2003JChPh.11912360M,2009JPCA..11310189L,2014JChPh.141p4324G}. Based on these studies, we estimate that for high collision energies considered in this study at current level of theory, the intersystem crossing effects are not likely to significantly affect the final results.
Third, we note that reactive processes were not considered in this study. \citet{2012JPCA..116...64Y} calculated the branching fractions for nonreactive scattering to be 0.984, as compared to the oxygen-atom abstraction reaction $(4 \times 10^{-6})$ and O atom exchange (0.016), with the uncertainties of about 30\% cited for all processes. Thus, we estimate that neglecting nonreactive channels contributes to no more than 2\% uncertainties in the reported cross sections.

\subsection{Implications on planetary aeronomy and escape of hot oxygen from Mars}


Current estimates of photochemical oxygen escape rates at Mars vary between about 10$^{24}$ s$^{-1}$ to 10$^{26}$ s$^{-1}$. While the numerical methods used between the studies are different (commonly Monte Carlo simulations vs two-stream solutions), O+CO$_2$ cross sections were identified as the key quantity responsible for the O escape rate variations \citep{2017JGRA..122.1102C,2018Icar..315..146J}. 

The O($^3P$)+CO$_2$ cross sections computed in this work are between 30\% and 50\% smaller than the values obtained by mass-scaling or semi-empirical methods that have been commonly used in interpretations of Mars orbiter data and simulations of photochemical oxygen escape\citep{}. Specifically, our \textit{elastic} cross sections are about $1.5\times10^{-14}$ cm$^2$ at 1 eV collision energy and about $8.5\times10^{-15}$ cm$^2$ at 5 eV. In comparison, the widely used elastic cross sections of \citet{2014ApJ...790...98L} evaluate to $2.24\times10^{-14}$ cm$^2$ and $1.97\times10^{-14}$ cm$^2$ for the two collision energies, respectively. 
In addition, the cross sections obtained by mass-scaling themselves scale with collision energy in the same way as the system on which they were based (typically O-O cross sections constructed for energy transfer in the terrestrial atmosphere by \citet{2000JGR...10524899K}), or are assumed constant over the entire collision energy range \citep{}. Similar arguments extend to differential cross sections: they either resemble the original system, or do not include angular anisotropy. 

We can estimate the impact of newly computed cross sections presented here on the hot oxygen escape flux at Mars.  \citet{2017JGRA..122.1102C} recently constructed a simple but robust analytic model of hot oxygen escape from Mars that scales well over a large range of solar EUV irradiances. 
Under the assumptions from their model, the photochemical oxygen escape flux from Mars is given by
\begin{equation}
  F_\mathrm{esc,O} \approx 0.1 \times I_\mathrm{CO2} (1 \; \mathrm{AU}) / \sigma^\mathrm{back}_{\mathrm{O-CO2}} \; ,
\end{equation}
where $I_\mathrm{CO2} (1 \; \mathrm{AU})$ is the photoionization frequency for CO$_2$ at 1 AU distance from the Sun and $\sigma^\mathrm{back}_{\mathrm{O-CO2}}$ is the total backscattering cross section for O+CO$_2$ collisions. 
Global escape rate can be approximated as $Q_\mathrm{esc} = 2 \pi R_\mathrm{Mars} F_\mathrm{esc,O}$, where it is assumed that the escape takes place from an ideal hemisphere under solar illumination.

\begin{table}
	\caption{Global Photochemical Oxygen Escape Rates at Mars.}
	\centering
	\begin{tabular}{lrr rrr}
		\hline
		$Q_\mathrm{esc}$  [s$^{-1}$]$^a$  & \multicolumn{2}{c}{Collision energy} & \multicolumn{3}{c}{Literature$^\mathrm{c}$} \\
                                  & 2.5 eV                   & 3.5 eV  & C17  &  L17  & FH18 \\  \hline
		solar maximum   & $20^\mathrm{a}$  & 16     & 9  &       & 4    \\
		solar minimum    & 7                           & 5      & 3  &       & 0.7   \\
		solar moderate   & 13                          & 10     & 6  &  4.3  &       \\  \hline
		$\sigma^\mathrm{back}_{\mathrm{O-CO2}}$ [$10^{-16}$ cm$^2$]  & 6.27 & 8.05 & 13 &  \\  
		\hline
		\multicolumn{6}{l}{$^\mathrm{a}Q_\mathrm{esc}$ in units of $10^{25}$ s$^{-1}$.} \\
		\multicolumn{6}{l}{$^\mathrm{c}$C17=\citet{2017JGRA..122.1102C}, L17=\citet{2017JGRA..122.3815L},} \\
		\multicolumn{6}{l}{\hspace{0.4em}FH18=\citet{2018Icar..300..411F}.} \\
	\end{tabular}
	\label{table_escapeflux}
\end{table}

In Table \ref{table_escapeflux} we compare global O escape rates calculated using the total backscattering cross sections at 2.5 eV and 3.5 eV constructed in this paper (see Fig. \ref{fig:DCS_selectedE}) with representative values given in the literature. A more complete list of estimated O escape rates with references can be found in \citet{2017JGRA..122.1102C}. The solar activity (irradiance) index $F_{10.7}$ is used to define solar maximum ($F_{10.7}=200$), minimum ($F_{10.7}=70$), and moderate activity ($F_{10.7}=130$). 

Our estimated photochemical O escape rates are larger than in \citet{2017JGRA..122.1102C} by about a factor of two and significantly larger than \citet{2018Icar..300..411F} for all levels of solar activity, with the differences being greater at 2.5 eV than at 3.5 eV. 
Consequently, we can expect photochemical hot O escape rates from Mars estimated using mass-scaling O+CO$_2$ cross sections would be up to two times greater if the cross sections presented in this work would be used instead. 
For example, we estimate that the escape rates constructed by \citet{2017JGRA..122.1102C} would increase to $\approx8.5 \times10^{25}$ s$^{-1}$. In comparison, the hot O escape rates derived from MAVEN observations of H$^+$ and O$^+$ ions picked up by the solar wind (i.e., pickup ions) have a mean value of $9 \times 10^{25}$ s$^{-1}$ with a factor of two variability \citep{2018JGRE..123.1192R}. 
Thus, a significantly better agreement between the two rates derived from measurements performed using different instruments onboard MAVEN orbiter could be reached if the cross sections presented here were used to model O+CO$_2$ collisions and their impact on attenuation of the hot O escape flux.

\section*{Acknowledgements}

MG and KZ have been partially funded by NASA, grant \#17-MDAP17\_2-0152. The authors are grateful to J. Fox, Y. Lee, and B. Jakosky for helpful suggestions and discussion, and to M.-P. de Lara-Castells for sharing with us a digital version of their potential energy surface points. An expert reviewer is acknowledged for providing helpful comments and suggestions that improved the quality of the article.




\bibliographystyle{mnras}
\bibliography{manuscript_arxiv_v3} 




%
%


\bsp	
\label{lastpage}
\end{document}